%% file: paper_main.tex
\documentclass[electronic]{vgtc}             

\graphicspath{{figures/}{pictures/}{images/}{./}} 

\usepackage{times}                     

\usepackage{tabu}                      
\usepackage{booktabs}                  
\usepackage{lipsum}                    
\usepackage{mwe}                       
\usepackage{wrapfig}
\usepackage{mathptmx}                  

\onlineid{1100}

\vgtccategory{System or Tool}

\vgtcinsertpkg

\title{ParaView-MCP: An Autonomous Visualization Agent with Direct Tool Use}

\author{Shusen Liu \and Haichao Miao \and Peer-Timo Bremer}
\vspace{-12mm}
\affiliation{\scriptsize CASC, Computing, Lawrence Livermore National Laboratory\\
\{liu42, miao1, bremer5\}@llnl.gov}

\teaser{
  \centering
  \includegraphics[width=1.0\linewidth]{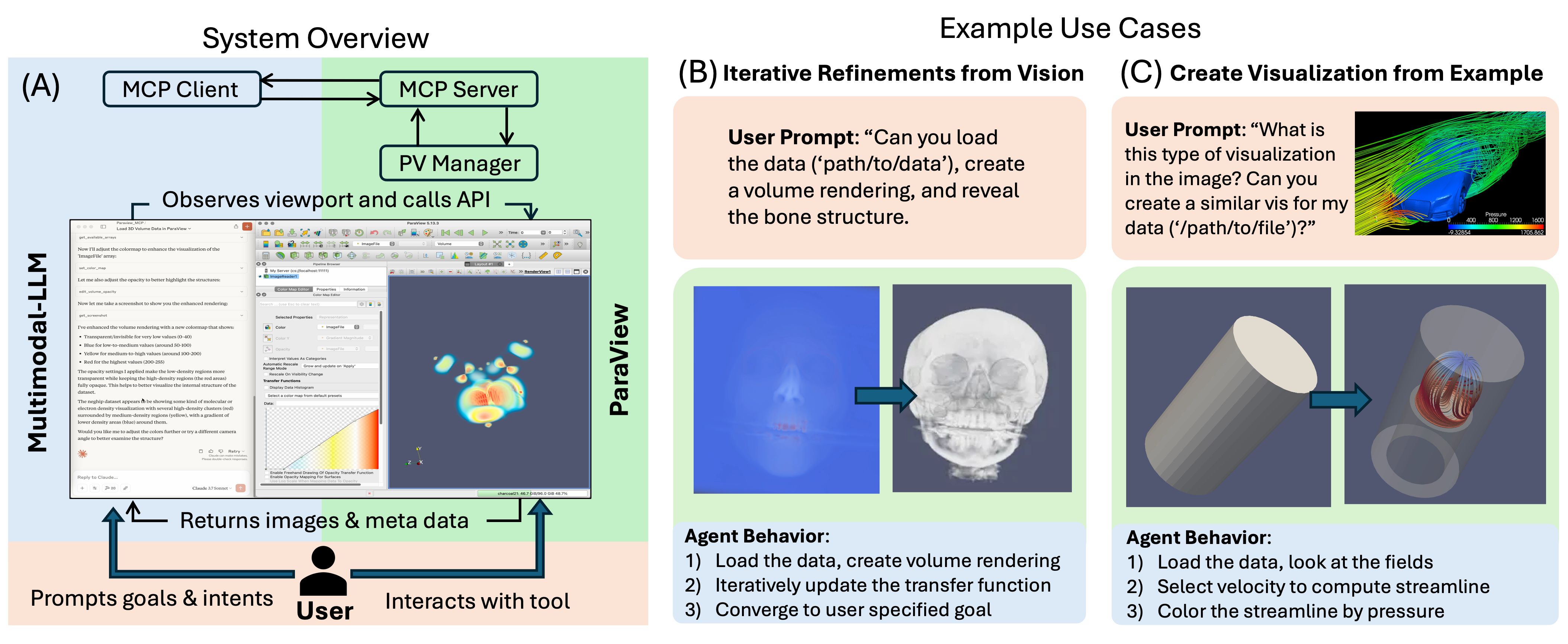}
  \vspace{-6mm}
  \caption{ParaView-MCP brings together the users, the tool, and the agent into a collaborative environment, where users can either express goals via the agent to achieve the visualization objective or directly interact with the ParaView GUI. (A) System Overview: We encapsulated ParaView's API via the ParaView Manager and developed an MCP Server to allow a multimodal large language model (MLLM) to directly control ParaView. By combining the MLLM's and ParaView's capabilities, we show novel usages that were not possible before with these tools alone, as shown in the examples (B, C). 
  }
  \label{fig:teaser}
}

\input{body/abstract}


\begin{document}


\input{body/intro}
\input{body/related}

\input{body/method}
\input{body/implementation}
\input{body/results}
\input{body/discussion}




\acknowledgments{
This work was performed under the auspices of the U.S. Department of Energy by Lawrence Livermore National Laboratory under Contract DE-AC52-07NA27344 (Review and released under LLNL-CONF-2005387). The work was partially supported by the LLNL-LDRD (23-ERD-029, 23-SI-003), and DOE ECRP (SCW1885).
}

\bibliographystyle{abbrv-doi}

\bibliography{refs}
\end{document}

%% file: body/abstract.tex
\abstract{
While powerful and well-established, tools like ParaView present a steep learning curve that discourages many potential users. 
This work introduces ParaView-MCP, an autonomous agent that integrates modern multimodal large language models (MLLMs) with ParaView to not only lower the barrier to entry but also augment ParaView with intelligent decision support. By leveraging the state-of-the-art reasoning, command execution, and vision capabilities of MLLMs, ParaView-MCP enables users to interact with ParaView through natural language and visual inputs. Specifically, our system adopted the Model Context Protocol (MCP) - a standardized interface for model-application communication - that facilitates direct interaction between MLLMs with ParaView's Python API to allow seamless information exchange between the user, the language model, and the visualization tool itself. 
Furthermore, by implementing a visual feedback mechanism that allows the agent to observe the viewport, we unlock a range of new capabilities, including recreating visualizations from examples, closed-loop visualization parameter updates based on user-defined goals, and even cross-application collaboration involving multiple tools. 
Broadly, we believe such an agent-driven visualization paradigm can profoundly change the way we interact with visualization tools. We expect a significant uptake in the development of such visualization tools, in both visualization research and industry. 
}

%% file: body/intro.tex
\firstsection{Introduction}
\maketitle

Scientific Visualization (SciVis) plays a crucial role in modern data-driven research and engineering workflows. Tools such as ParaView \cite{ahrens2005paraview} are widely recognized for their powerful capabilities, scalability, and flexibility across a broad range of application domains. 
While the intent and desire to use these tools are typically clear, many domain scientists struggle to incorporate them into their workflows effectively, often left with utilizing suboptimal visualization methods. 
A key barrier to adoption lies in the opportunity cost and steep learning curve associated with mastering these tools. They often cannot justify the time and effort required to learn a new tool, especially when their focus lies elsewhere. 
As a result, powerful visualization platforms like ParaView remain underutilized, despite being freely available and well-supported. 
Anecdotally, many application scientists express sentiments such as: “I plan to learn to use it, but I never get around to it.” The threshold for entry will be significantly lowered if someone is available to guide them hands-on, using their own data and use cases.

With the advent of modern multimodal large language models (MLLMs), new opportunities arise for improving tool usability. These models not only possess broad general knowledge but can also issue structured commands and follow complex, goal-oriented instructions. This opens up exciting possibilities for rethinking how users interact with software, not through complex and hard to master traditional user interfaces, but via conversational, intelligent agents that can understand user intent and act on their behalf.

In this work, we present ParaView-MCP, an \emph{autonomous visualization agent} that integrates a state-of-the-art MLLM with the ParaView platform. Our system leverages the Model Context Protocol (MCP) \cite{anthropic2024mcp}, a model-application exchange standard, and ParaView’s Python API to establish a direct, bidirectional communication pipeline between the user, the language model, and the visualization tool. This integration allows for seamless command execution, visualization manipulation through natural language, and context-aware interactions.
Through this setup, users can perform tasks that previously required extensive training and/or manual effort. For instance, the agent can recreate visualizations from examples, automatically update plots based on user-defined goals, or even coordinate across multiple tools in a collaborative analysis scenario. These capabilities illustrate how AI-driven agents can significantly extend the usability and expressiveness of existing software systems.

Ultimately, ParaView-MCP aims to lower the barrier to entry for high-quality scientific visualization, transforming how users access, control, and benefit from complex visualization tools. 
This paper outlines the system's design, implementation, and capabilities and highlights a set of case studies that demonstrate its potential in real-world scenarios. 
By facilitating direct tool use via MCP‑enabled multimodal agents, our work unites three previously disparate threads that make decades of SciVis algorithms embedded in ParaView more accessible to domain users via conversational interfaces that democratize the visualization of complex datasets.

%% file: body/related.tex
\section{Related Work}
\textit{ParaView-MCP} sits at the intersection of traditional SciVis techniques, natural‑language and conversational interfaces, and the recent wave of AI-assisted visualization systems. On the traditional SciVis side, volume rendering with transfer‑function design and iso‑surface extraction ~\cite{levoy1988surfaces} remain indispensable for various applications, yet domain experts still report that tuning transfer functions or iso‑values is non‑intuitive and time‑consuming~\cite{ljung2016transfer}. 
ParaView and VisIt expose these capabilities through rich data‑flow pipelines, but the GUI and scripting APIs require considerable expertise. 

Making visualization tools accessible through natural language has been a long‑standing goal. 
Early systems such as \textit{DataTone} handled linguistic ambiguity via interactive widgets~\cite{gao2015datatone}; \textit{Eviza} enabled follow‑up Q\&A and incremental view refinement in conversational form~\cite{setlur2016eviza}. 
Subsequent toolkits (NL4DV, FlowSense, etc.) broadened the mapping from natural language to Vega‑Lite or dataflow specifications \cite{narechania2020nl4dv, wang2022towards}. 
These works demonstrate that natural language lowers the entry threshold, yet they typically focus on 2D \textit{InfoVis} that output visualization specifications rather than controlling a running \textit{SciVis} application.

With the recent advancement, AI-centric visualization techniques \cite{wu2021ai4vis, ye2024generative} have grown dramatically and spread through most subfields of visualization, e.g., visualization education \cite{kim2024chatgpt}, visualization design/recommendation \cite{hu2019vizml, wang2023llm4vis, wang2024dracogpt, berger2024visualization, xie2024haichart, ouyang2025nvagent}, compression \cite{lu2021compressive}, and super-resolution \cite{han2020ssr}. 
Moreover, the visual‑understanding capabilities of modern MLLMs have sparked studies for evaluating how well these models interpret visualization~\cite{yang2023dawn, bendeck2025gpt4, li2025recall, vazquez2024llmsreadyvisualization, xu2024svg}.  
Although results reveal notable shortcomings, the overall trend points to rapid improvement, providing growing confidence that MLLMs can supply meaningful visual reasoning. 

Another significant change comes from research on AI agents \cite{plaat2025agentic}, where autonomous decisions are carried out based on human preferences.
Visualization researchers have previously explored agent‑like systems: \textit{VisComplete} auto-completes visualization pipelines using histories \cite{koop2008viscomplete};  
\textit{ChatVis} iteratively refined ParaView Python scripts with an LLM and error‑feedback loop~\cite{mallick2024chatvis}; The \textit{MatplotAgent} \cite{yang2024matplotagent} work assesses various models' ability to generate matplotlib code for general visualization tasks; 
\textit{AVA} optimized volume‑rendering parameters through closed‑loop optimization~\cite{liu2024ava}.
While compelling, these systems stop short of direct in‑situ control of an existing tool’s interface during an interactive exploration and analysis session.
Recent frameworks such as \textit{HuggingGPT} illustrate already how an LLM can call external tools to solve collaborative tasks across tools~\cite{shen2023hugginggptsolvingaitasks}. Anthropic \cite{anthropic2024mcp} and OpenAI \cite{openai_mcp_2025} have generalized this idea with the MCP that allows tools to expose functions to language agents.

%% file: body/method.tex
\section{ParaView-MCP}
Here we outline the design and implementation of the ParaView-MCP system. We first examine the two dominant existing design paradigms for tool use and then discuss the proposed hybrid approach, used for ParaView-MCP. Furthermore, we discuss the system implementation and provide the rationales and design choices. 

\subsection{Autonomous Visualization Agent Design}

\noindent\paragraph{Command-Line / API Interface}
Many tools can be controlled through direct code generation, such as those used in ChatVis \cite{mallick2024chatvis}.
While this method is relatively simple to implement and effective for experienced users, it is inherently limiting, as it does not support collaborative environments where both the agent and the user can interactively operate the tool in real-time, restricting its applicability for complex human-tool-agent interactions. Furthermore, the lack of a GUI limits the range of possible applications. 

\noindent\paragraph{Pure Vision-based GUI Control}
Vision-based computer use, where the agent interacts with the GUI as a human user would (controlling mouse and keyboard), offers a general solution applicable to a wide range of tools. However, current implementations, such as OpenAI’s Operator \cite{openai2025operator} or Anthropic's Computer Use \cite{anthropic2024computeruse}, lack the reliability and speed necessary for effective interactive operation. The slow response times and challenges in navigating complex interfaces prevent this approach from being viable for real-time collaborative workflows.

\paragraph{Our Proposed Hybrid Approach: API Control + Focused Visual Understanding}
Our approach leverages the strengths of direct API integration while focusing the model's vision capability where it is needed most, i.e., interpreting the visualization in the viewport. By utilizing ParaView’s existing Python API, the agent can directly manipulate the tool without requiring any alteration to its software architecture. 
This hybrid design enables interactive operation where both the user and the agent can collaboratively manipulate the visualization and observe the updates. 
Moreover, the visual feedback further reduces potential hallucination issues by grounding the agent’s decisions in the visualization results. While natural language interaction is a key component, the ability to iteratively refine actions based on visualization results ensures accuracy and precision, making the system robust and accessible for both expert and non-expert users.

%% file: body/implementation.tex
\subsection{System Implementation}
\label{sec:implementation}
The hybrid scheme discussed above is achieved by creating and exposing a structured tool abstraction of ParaView to the language-model via an MCP server. 

\paragraph{ParaView Control Via Multi-Client PvServer}
ParaView offers a comprehensive Python API, \texttt{paraview.simple}, which enables programmatic access to nearly all aspects of the tool—from data loading and processing to visualization workflows and rendering. 
To allow control of ParaView from both the GUI interface as well as the agent while maintaining interactivity, we utilize ParaView’s \texttt{pvserver} in multi-client mode. In this setup, both the user (through the ParaView GUI) and the autonomous agent (through ParaView-MCP Server connected to the same \texttt{pvserver}) can issue commands and observe updates in a shared session. This enables seamless, collaborative control over the same visualization state without sacrificing user agency or GUI responsiveness. 
\paragraph{ParaView-MCP Server}
At the heart of our system lies the ParaView-MCP server, which bridges the language model and the visualization tool. This server is built on the \textit{Model Context Protocol (MCP)} \cite{anthropic2024mcp}, a standardized framework designed to provide LLMs with structured, contextual access to external tools.
MCP allows the tool usage to be LLM-agnostic, compatible with different model providers or architectures, while preserving a consistent and interpretable tool interface. This abstraction makes the system extensible and adaptable as language model's capabilities evolve.
However, simply exposing the low-level ParaView API directly as functions/tools for the MCP server is not a viable solution: 1) a large number of functions may make selecting the correct one more challenging; 2) the protocol is not designed to work with arbitrary function returns that are native to ParaView. 
The abstraction needed by the MCP Server defines the range of possible ParaView-MCP behavior, as a result, it is the most critical design choice for this work. 
There are essentially two main avenues to approach this challenge. First, we can develop a set of hard-coded, predefined tools, i.e., discrete operations corresponding to common ParaView tasks such as loading datasets, applying filters, adjusting views, or exporting results. 
Alternatively, we can leverage the code generation capabilities of advanced LLMs to produce custom code snippets on the fly, and then execute them on demand. This approach provides increased flexibility by translating high-level language instruction directly into lower-level APIs calls, however, it also introduces challenges in safety, correctness, and repeatability. Due to these potential issues, we adopt the former approach, where each function is carefully crafted to support a key usage. 
Based on such an abstraction, the ParaView-MCP server exposes a curated set of ParaView functionalities to the language model, along with rich metadata describing each function’s purpose, input/output types, and parameter constraints. This interface acts as both a documentation and a command execution layer, ensuring the language model can reason about and safely invoke available functionality.

\paragraph{ParaView Manager}
The \textit{ParaView Manager} encapsulates what functionalities the MCP Server can access. 
This ensures reliable execution, avoids unsafe code generation, and supports repeatability—important for scientific reproducibility and debugging. 
The LLM will handle the mapping from language to the set of appropriate tools, validate returns, and append information to the LLM context for future reference.
One persistent challenge we encounter when designing this encapsulation is tracking the active object and visualization state, Ideally, we prefer the abstraction layer to be stateless. To achieve that, we expose specific functions to query existing ParaView \textit{sources} by type or name, select active \textit{sources}, and determine whether the operation can be applied to the source.
For the future iterations of ParaView-MCP, dynamic tool synthesis may offer additional flexibility, but for now, we prioritize robustness and traceability.
Moreover, the ParaView Manager also facilitates the query of the viewport during the interaction, which facilitates vision-grounded reasoning that sets the proposed approach apart from many existing generative visualization approaches that focus on image and code generation rather than visual assessment or closed-loop iteration. 
Lastly, we plan to open-source the implementation of ParaView-MCP.\footnote{
\textit{At the time of submission, the Paraview-MCP repository is under our organization's Information Management review.}
}

%% file: body/results.tex
\section{Use Cases}
In this section, we illustrate potential use cases for ParaView-MCP.
While these examples highlight interesting functionalities, the flexibility of the system enables a wide range of possibilities beyond what can be covered here.
We also refer the reader to the supplementary video \footnote{See the video demo at: \url{https://youtu.be/GvcBnAcIXp4} } for the full interaction of these use cases. 

\subsection{Language to Result with Vision-Guided Refinement}

\begin{wrapfigure}{r}{0.5\linewidth}  
    \centering
    \vspace{-3mm}
    \includegraphics[width=\linewidth]{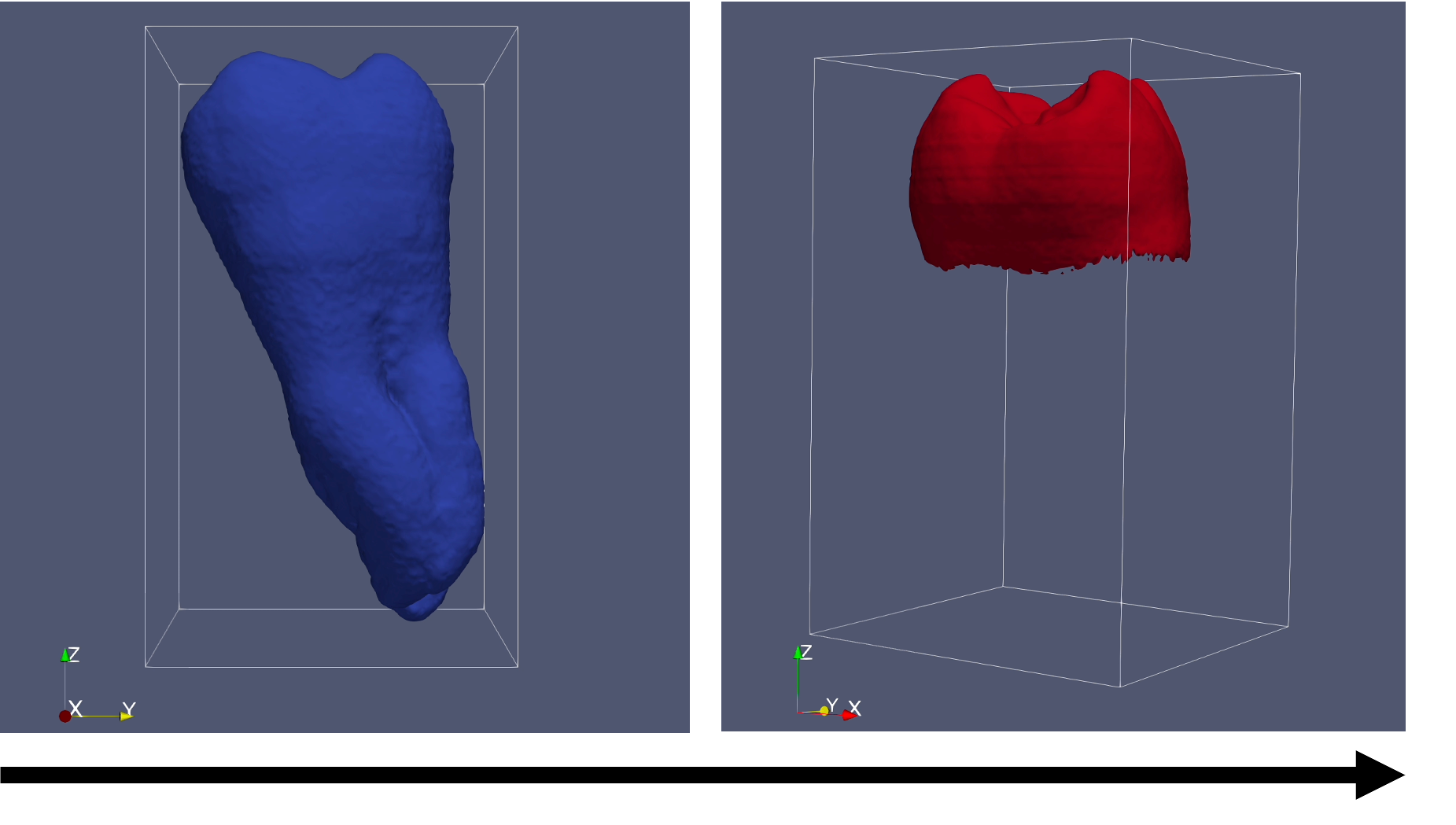}
    \vspace{-8mm}                     
    \caption{
    The agent automatically identifies the iso-value that leads to half of the original surface area.}
    \label{fig:isosurface_example}
    \vspace{-2mm} 
\end{wrapfigure}

ParaView-MCP translates users' goals and intentions into MCP function calls that control the ParaView GUI.
The user can directly ask for specific features, like creating an isosurface or volume rendering of their volumetric data. 
More interestingly, the agent can pursue user-defined objectives iteratively. 
As illustrated in Figure \ref{fig:isosurface_example}, a user could specify the goal to identify an isovalue that generates a mesh surface that is roughly half of the original one. ParaView-MCP will then iteratively adjust the iso values until this goal is met.

\begin{figure}[!htbp]
    \centering
    \includegraphics[width=1.0\linewidth]{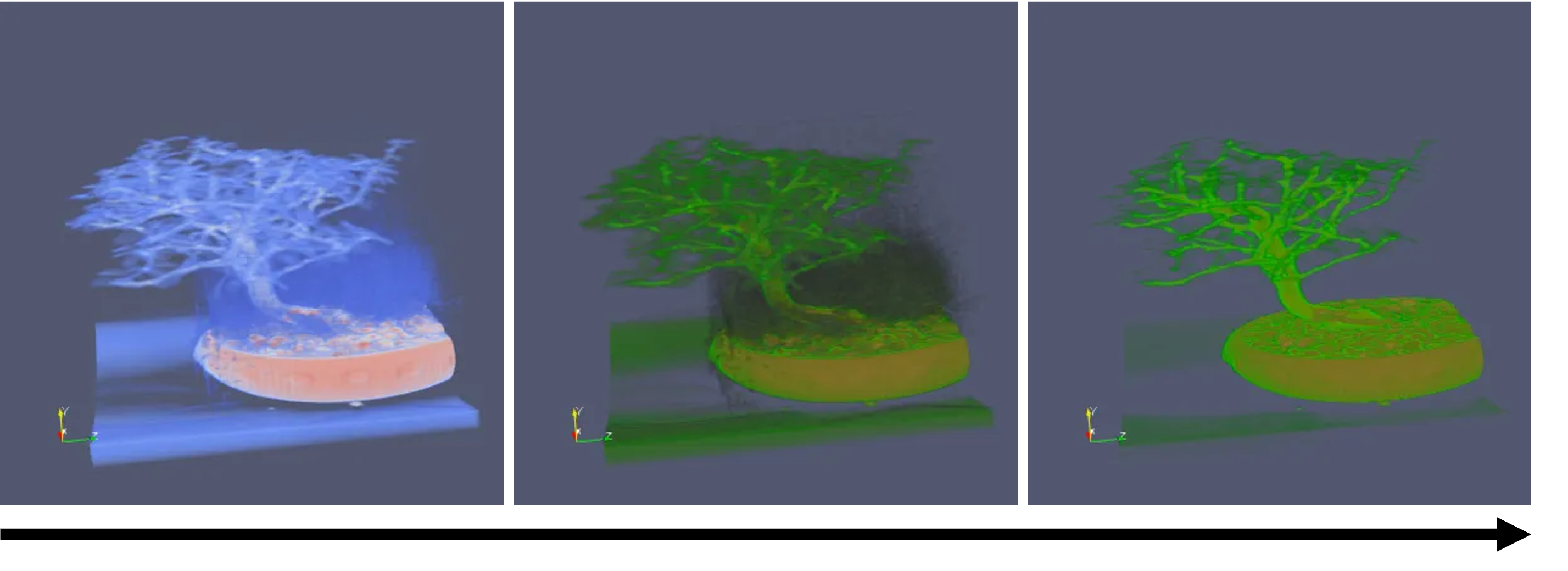}
    \vspace{-8mm}
    \caption{Iterative transfer function design driven by a vision-based feedback loop. The agent starts with the default colormap (\emph{Cold-to-Warm}) to assess the value distribution, and based on user instruction (\emph{green tree, brown base}), designs and refines the colormap by visually examining each rendering output.}
    \label{fig:TF_design_example}
    \vspace{-4mm}    
\end{figure}

Moreover, by leveraging vision-based feedback and function return values, the system enables iterative refinement to achieve user-defined objectives. 
As shown in Figure \ref{fig:TF_design_example}, the user specifies the target color for both tree and base, the agent follows the instruction, recognizes the structure in the image, and then tries to assign the correct color to the right object, iteratively.
Similarly, in Figure \ref{fig:teaser}(b), the agent successfully designs the transfer function that highlights the bone structure.
Transfer function design process deals with a large action space, which could benefit from a gradient-based optimization approach such as proposed by Jeong et al. \cite{jeong2024text}. However, their approach requires not only a fully differentiable volume renderer but also the CLIP \cite{radford2021learning} scoring model that has a weaker vision capability compared to the frontier MLLMs (e.g., GPT, Claud).

\begin{figure*}
    \centering
    \includegraphics[width=\linewidth]{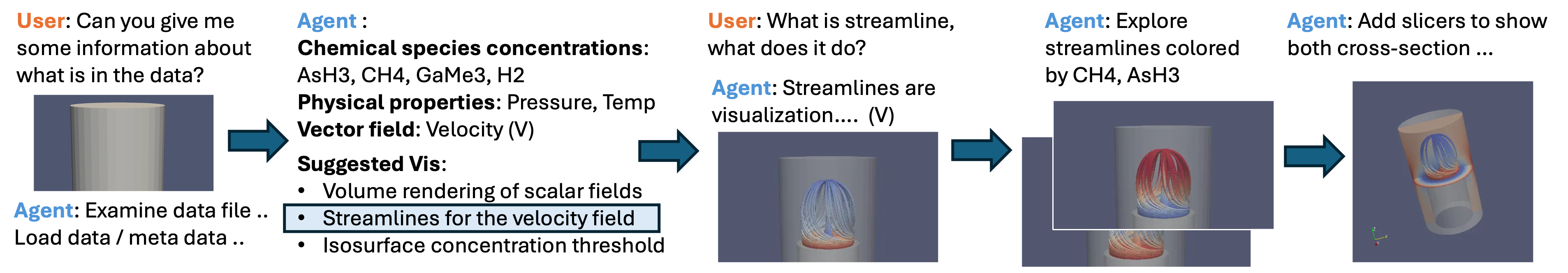}
    \vspace{-6mm}
    \caption{
    An illustration of the co-exploration use case, where the ParaView-MCP works with a novice user by suggesting a list of possible visualization methods, explaining specific techniques (e.g., streamline), and guiding the overall exploration process.
    }    
    \label{fig:co_exploration}
\end{figure*}

\subsection{Visualization by Example and Co-Exploration}
The multimodal capabilities of the MLLM enable users to provide visual references alongside language instructions. For instance, a user could request the agent to replicate a visualization based on a visual example, as illustrated in Figure \ref{fig:teaser} (c). This ability to combine visual and linguistic inputs enhances the agent’s understanding and facilitates the recreation of potentially complex visualization workflows.
Additionally, the conversational interface of ParaView-MCP facilitates collaborative exploration by allowing novice users to ask questions about their data and potential visualization solutions. As illustrated in Figure \ref{fig:co_exploration}, the user start by asking the agent to help explain the loaded data, then the agent not only provide a detailed breakdown of the content of the data but also suggest and explain suitable visualization approaches, and work with the user to explore the different type of data/field together.

\subsection{Multi-Tool Interaction and Collaboration}
\begin{figure}[!htbp]
    \centering
    \includegraphics[width=1.0\linewidth]{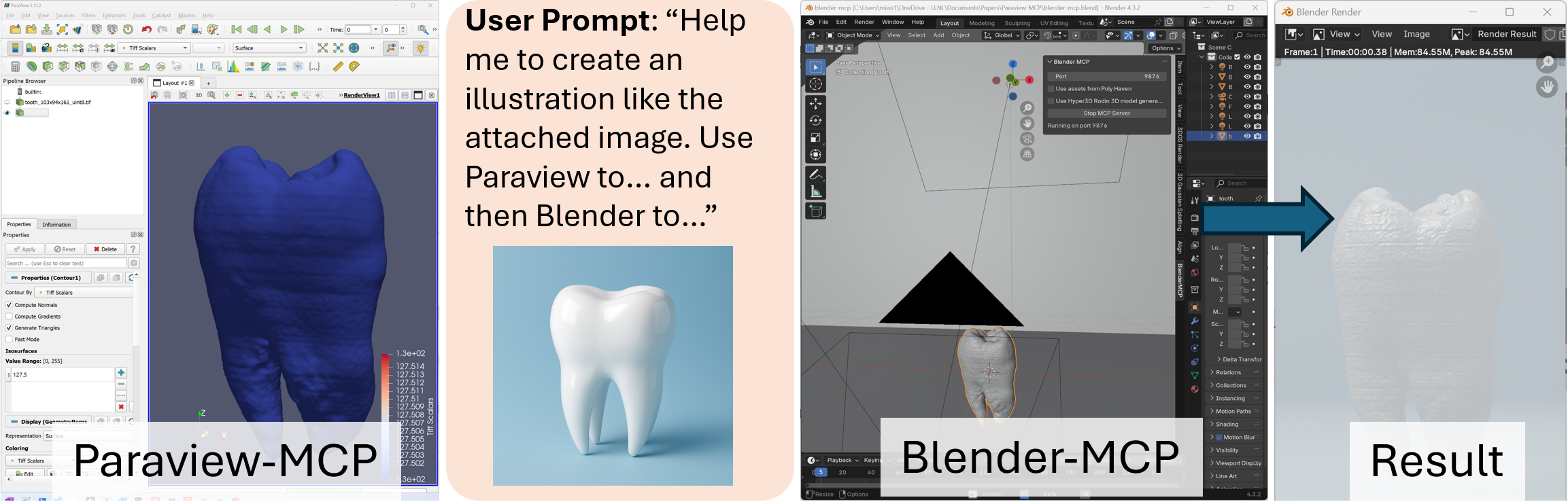}
    \vspace{-6mm}
    \caption{A multi-tool collaboration example use case, where Paraview-MCP and Blender-MCP \cite{ahujasid_blender_mcp} work together to create an illustration from a volumetric dataset of a tooth. Per user prompt, ParaView-MCP extracts the iso contour and then renders the surface in Blender-MCP based on the example image the user provided.}
    \label{fig:multi_tool}
     \vspace{-4mm}
\end{figure}

Visualization tasks often utilize not one but multiple tools. Here we demonstrate a use case where, with the help of a second MCP server for another tool, an illustrative visualization is generated from a volume dataset. 
Multiple MCP servers can connect to the same MCP client (e.g., the Claude Desktop App), enabling seamless integration across different tools. As shown in Figure \ref{fig:multi_tool}, the agent can generate an iso-contour using ParaView-MCP and subsequently send the mesh into Blender via the \emph{Blender MCP server} \cite{ahujasid_blender_mcp}. This interoperability allows users to leverage the strengths of multiple tools in a unified workflow, facilitating advanced cross-platform analysis and visualization tasks.


\section{Preliminary Domain Expert Feedback}
To obtain rapid, formative insights on the utility and research relevance of ParaView-MCP, we invited two domain experts to review the described prototype. 
As intended, neither had used ParaView before, but both actively work with CT data and had heard of ParaView. E1 is a research scientist in non-destructive testing who specializes in the acquisition and analysis of CT scans. E2 is a program manager who oversees a portfolio of research projects in additive manufacturing and material characterization. We demonstrated the tool using the transfer function design and isosurface selection examples, and reported their informal feedback below. 
E1 said \textit{"the fact that it directly controls ParaView is incredible"}. E2 commented that she not only sees the value proposition for novice users, but also experienced ParaView users who can \textit{"use this tool and let it automatically explore a bunch of datasets"}. When demonstrating the transfer function design use case to visualize the bonsai tree like a \textit{real tree}, E1 noted that after the initial prompt, the agent chose an inverted color map, since lower density values, corresponding to leaves, are mapped to brown and higher density values, corresponding to the tree trunk, are mapped to green. With a second prompt \textit{"could you please render the lower density that represents leaves in a different color than the wood and the pot?"} the agent was able to color the leaves in a different color. He found the text explanation from the agent while inspecting the result in ParaView in particularly useful to understand the agent's reasoning and also to learn about ParaView usage. E2 also made an insightful suggestion that the agent should be able to understand all the actions that the user conducted in the ParaView GUI, beyond only observing the viewport, which we plan to support. Both users agreed that this tool lowers the barrier of entry to a complex visualization tool. 

%% file: body/discussion.tex
\section{Discussion}

\paragraph{Agentic Paradigm for Visualization}
Since the advent of visualization for medical and scientific use cases, the visualization research community has developed a vast number of custom techniques/tools for ever-changing domain applications. Yet, the actual adoption of these custom visualization methods remains low. ParaView-MCP does not propose a new visualization method, but a visualization agent with a direct tool use paradigm. This paradigm shift could address the ``last mile'' problem for the adoption of existing visualization methods by significantly lowering the barrier of entry. Specifically, ParaView-MCP can directly control ParaView, observe its viewport, and interact with the user via natural language. To the best of our knowledge, this is the first implementation of the Model Context Protocol (MCP) for a complex SciVis tool, demonstrating its applicability and effectiveness in handling sophisticated visualization workflows. We believe autonomous visualization agents would significantly alter how we interact with tools in the future. We hope to encourage the creation of similar autonomous visualization agents by demonstrating their viability and strengths. At the same time, we also want to discuss their limitations, provide guidance on the design, and cover lessons learned. 

\paragraph{Challenges and Lessons Learned}
In the current work, ParaView-MCP utilizes Claude's Sonnet 3.7, which is Anthropic's most intelligent model. Still, we observed the limitation of this model when controlling ParaView. For example, as noted by E1, it inverted the color map for the bonsai tree even though it should have had general knowledge about the density values of leaves vs. wood. On some occasions, it would oscillate between several wrong color maps without ever reaching the correct solution, even though it observes the changes it made via the viewport. Besides the inherent stochasticity of LLM, erroneous behaviors like that also spark the question of how intelligent these models are, if they can conduct highly complex tasks yet fail at seemingly common knowledge. Because of cases like this, the human-in-the-loop will remain ultimately crucial in high-consequence workflows. Specifically, the visualization agent paradigm introduced here advocates a collaborative process, where the agent supports the human expert, such as tedious and laborious tasks, with direct tool use, but the user ultimately remains in charge.  

For the important lessons learned. First, from our initial experiment, the speed of interaction, seemingly trivial, has a big impact on the user experience.
Its importance becomes painfully obvious when we experiment with pure vision-based tool use, i.e., Claude Computer Use API ~\cite{anthropic2024computeruse}. Besides limited navigation accuracy, a single interaction step (mouse click/movement, then vision feedback from the entire screen) takes somewhere between 10-20 seconds, and opening a menu sometimes will require 3 steps.
As a result, we instead focus on designing tool-use strictly through the ParaView API, which is multiple magnitudes faster. 
Besides interactivity, as discussed in detail in Section \ref{sec:implementation}, creating a useful agent goes beyond exposing the API and viewport to the MLLM. Significant considerations have to be made to ensure consistent and predictable behavior of the agent, hence, we iterate different abstraction designs for the ParaView Manager.

\paragraph{Future Directions}
With our open-source release of ParaView-MCP, we anticipate feedback on its capabilities and limitations to guide future feature implementation. Our immediate focus will be on enhancing batch processing abilities, particularly for opening datasets and creating/deleting sources, facilitating visualization tasks across multiple datasets. A reliable API interface forms the necessary foundation. The \emph{multi-client} option isn't optimal for agent/tool communication, being recently deprecated in ParaView. A more robust solution would be to develop a ParaView plug-in to directly inject commands into its Python console. Looking further ahead, combining MLLM agentic frameworks for high-level planning with dedicated optimizers (e.g., gradient-based TF design \cite{jeong2024text}) for lower-level tasks presents an intriguing direction, potentially delivering the best of both approaches while improving speed and efficiency.